\documentstyle[12pt,epsfig,wrapfig]{article}
\begin{document}
\begin{flushright} NIKHEF 96-023\\ hep-ph/9610289 \end{flushright}
\begin{center}
{\large \bf
Quark masses in Qiu's factorization procedure\footnote{Contribution to the
proceedings of the 12th Int.\ Symposium on High-Energy Spin Physics, Amsterdam,
Sept 10-14,1996}
\\}
\vspace{1mm}
\underline{D. Boer} and R.D. Tangerman
\\
{\small\it
NIKHEF, P.O.Box 41882, 1009 DB Amsterdam, The Netherlands
\\ }
\end{center}

\begin{center}
\begin{minipage}{130 mm}
\small
We discuss a higher-twist factorization procedure of the hadron tensor in deep inelastic scattering, proposed by Qiu, and extend it to include quark 
mass terms. Its property that the hard scattering parts are electromagnetic gauge invariant separately, is manifestly preserved. 
Using an auxiliary parton, a so-called spurion, to generate the quark mass 
terms, a simple parton model interpretation is also retained.
\end{minipage}
\end{center}

In Ref.~\cite{poli80} Politzer suggests a diagrammatical
approach to unravel the power corrections in $1/Q$ for a wide class of
hard scattering processes.
This method was fully employed for the deep inelastic scattering (DIS) process
by Ellis, Furma\'nski, and Petronzio (EFP) in Ref.~\cite{EFP}, in which
they calculate the twist-four corrections to the unpolarized structure
functions. Thereto they factorize the diagrams in hard, i.e., perturbative, and
soft, i.e., non-perturbative, parts. 
It was Qiu ~\cite{qiu90} who showed that, if one uses the so-called {\em special propagator\/}, i.e., the non-propagating part of the propagator, one can factorize such that the electromagnetic gauge invariance of the hard parts is manifest.
It also enables a clear parton model
interpretation of the higher twist terms, i.e., each order of power suppression means taking into account correlation functions with one parton more.

Both EFP and Qiu neglect quark masses. However, in some calculations of DIS structure functions they cannot be neglected, e.g.\ the case of $g_2$ of a free quark target at tree-level~\cite{jaff91a}. Our goal is to include them in such a
fashion that Qiu's result is unaffected. Since the quark masses arise from two sources, namely from the hard parts and from the use of the equations of motion in the soft parts, this is a non-trivial question.
 
\begin{wrapfigure}{R}{11cm}
\vbox{\hbox{\epsfig{figure=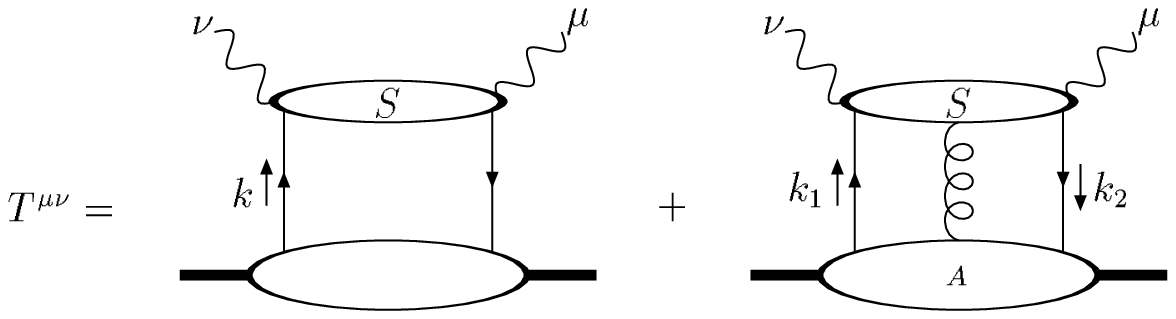,width=10cm}}
\vspace{4mm}
\parbox[t]{10cm}{\small Figure 1: Forward scattering amplitude.}}
\end{wrapfigure}
The starting point is the familiar diagrammatic expansion of the forward
scattering amplitude $T^{\mu\nu}$ (we consider only quarks of one flavor) ~\cite{EFP,efre84,AEL}
\begin{equation}
T^{\mu\nu}= \int d^{4}k \, {\hbox{Tr}} \left[ S^{\mu\nu} (k;m) \Gamma (k) \right]
+ \int d^{4}k_1  d^{4}k_2 \, {\hbox{Tr}} \left[ S^{\mu\nu}_{\alpha} (k_1,k_2;m)
\Gamma^{\alpha}_{A} (k_1,k_2) \right] + \ldots,
\label{expansion}
\end{equation}
keeping only the terms contributing up to order $1/Q$ (see Fig.~1).
Here, $S$ and $\Gamma$ are 
the hard and 
soft scattering parts, respectively.
We will not consider QCD corrections, so the hard parts consist of the forward
parton-photon scattering tree graphs which are 1PI in the $\gamma\gamma$-channel.

The next step is performing, in addition to the familiar collinear expansion, employed by EFP, a quark-mass expansion of the hard parts.
We define the Ward identity $\partial S^{\mu\nu} (k;m)/\partial m \equiv S^{\mu\nu}_{\hbox{\scriptsize{spur}}}
(k,k;m)$, where the right-hand side follows from the insertion of an auxiliary parton, a
zero-momentum (scalar) spurion~\cite{llew88} (denoted by a dashed line), which
couples (only) to
the fermions through the vertex $-i{\bf 1}_{ij}$. The
spurions will generate the mass contributions. 
One arrives at
\begin{eqnarray}
T^{\mu\nu} &=& \int dx \, {\hbox{Tr}} \left[ S^{\mu\nu} (x;0) \Gamma (x) \right]
+\int d x_1 d x_2 \, {\hbox{Tr}} \left[S^{\mu\nu}_{\alpha} (x_1,x_2;0)
{\omega^\alpha}_{\beta} \Gamma^{\beta}_{D} (x_1,x_2)\right]\nonumber\\
&&+ \int d x_1 d x_2 \, {\hbox{Tr}} \left[S^{\mu\nu}_{\hbox{\scriptsize{spur}}} (x_1,x_2;0)
\Gamma_{m} (x_1,x_2) \right]
+ \ldots, \label{efp}
\end{eqnarray}
where the $x$'s are the longitudinal momentum fractions of the partons and the projector
$\omega_{\alpha\beta}=g_{\alpha\beta}-p_{\alpha} n_{\beta}$.
The soft parts $\Gamma, \Gamma_D^{\beta}$ are given in~\cite{EFP,wij} and  $\Gamma_{m}$ reads
\begin{eqnarray}
\Gamma_{mij} (x_1,x_2) &\equiv&  \int \frac{d \lambda}{2\pi} \frac{d \eta}{2\pi}
e^{i\lambda x_1} e^{i\eta (x_2 -x_1)} \langle \, P,S|T\left[\overline{\psi}_j
(0) m
\psi_i(\lambda n)\right]| P,S \, \rangle .
\label{blobs}
\end{eqnarray}
\begin{wrapfigure}{r}{10cm}
\vbox{\hbox{\epsfig{figure=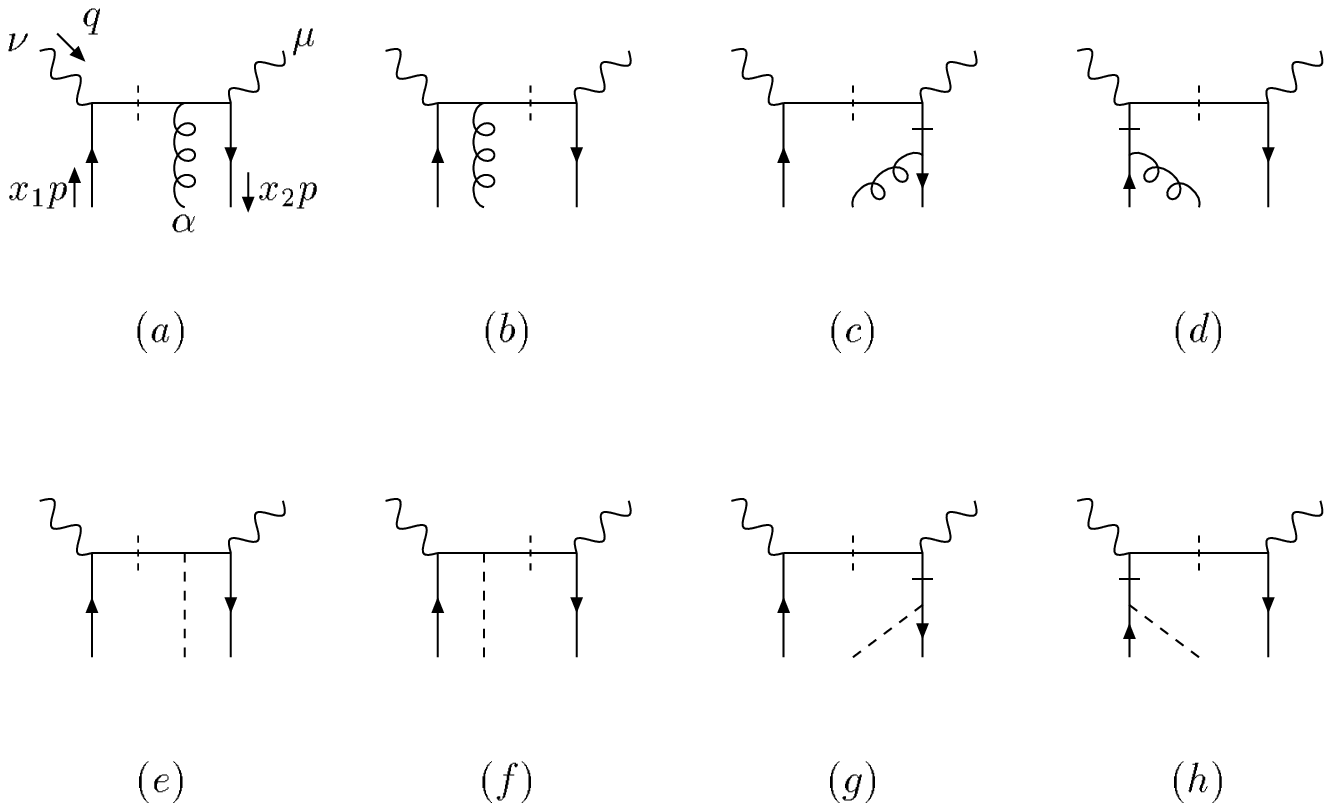,width=9cm}}
\parbox[t]{9cm}{\small Figure 2: Uncrossed twist-three gluonic [$(a)$-$(d)$]
and spurionic [$(e)$-$(h)$] diagrams.}}
\vspace{2mm}
\end{wrapfigure}
The first term in Eq.~(\ref{efp}) contributes to order $(1/Q)^0$ {\em and
higher},
whereas the other terms contribute to order $(1/Q)^1$ {\em and higher}.
The ingredient due to Qiu~\cite{qiu90} is basically the splitting of
the different terms into parts which contribute at a specific
order in $1/Q$. The sets of diagrams of a particular order in $1/Q$ constitute gauge invariant sets. These can only be identified by splitting off the mass contributions.

The
leading and sub-leading forward scattering amplitudes become \cite{wij}
\begin{eqnarray}
T^{\mu\nu}_{\hbox{\scriptsize{twist-2}}} &=&
\int dx \, {\hbox{Tr}} \left[ P_{-} S^{\mu\nu} (x;0) P_{+} \Gamma (x)
\right],\label{LO}\\
T^{\mu\nu}_{\hbox{\scriptsize{twist-3}}} &=&
\int dx_1 \, dx_2 \, {\hbox{Tr}} \left[ P_{-} H_{\alpha}^{\mu\nu}(x_1,x_2;0)
P_{+}
{{g_T}^\alpha}_{\beta} \Gamma^{\beta}_{D} (x_1, x_2) \right]\nonumber\\
&& +\int dx_1 \, dx_2 \, {\hbox{Tr}} \left[ P_{-}
H_{\hbox{\scriptsize{spur}}}^{\mu\nu}(x_1,x_2;0)
P_{+} \Gamma_{m} (x_1, x_2) \right].\label{NLO}
\end{eqnarray}
The modified hard gluonic and spurionic
parts, denoted by $H_{\alpha}^{\mu\nu}$ and $H_{\hbox{\scriptsize{spur}}}^{\mu\nu}$, are depicted in Fig.~2 after cutting the diagrams to arrive at the contributions to the hadron tensor.
The special propagator \mbox{$\mbox{$\not\! n\,$}/2x$}~\cite{kogu70,qiu90} is denoted by a slashed propagator (we use $p$ and $n$ to denote two orthonormal light-like vectors, where p is the dominant direction of the hadron momentum).
Diagrams (c) and (d) are due to Qiu. Diagrams (e)--(h) generate the mass terms.
The projectors are 
\mbox{$P_{+} = \mbox{$\not\! p\,$} \mbox{$\not\! n\,$}/2$},
\mbox{$P_{-} = \mbox{$\not\! n\,$}\mbox{$\not\! p\,$}/2 $}, 
\mbox{$g_T^{\rho\sigma} = g^{\rho\sigma}-p^\rho n^\sigma -n^\rho p^\sigma$}
and project on the `good' quark field, `bad' quark field and the `good' gluon field
respectively~\cite{kogu70}. 
They were not used in Ref.~\cite{qiu90}.
Due to the projections the terms (\ref{LO}) and (\ref{NLO}) are of one specific order in $1/Q$. 
In general, `twist-$t$' only contributes
at order $Q^{2-t}$. Also, $t$ always equals the number of partons connecting
the hard and soft parts. 

We found explicitly \cite{wij} that the (projected) discontinuities now satisfy electromagnetic gauge invariance, i.e., contraction with the photon momentum
gives zero. Other results of \cite{wij} are also in agreement with the standard EFP-type of
calculations~\cite{efre84,AEL}. 
The extension to higher twist is straightforward. 
The general algorithm can be given as follows. To obtain the
regular contributions to the order $Q^{2-t}$ DIS hadron tensor,
following Qiu~\cite{qiu90}, one has to (1) write down {\em all\/} possible forward
2-photon $t$-parton cut diagrams,
where partons can be quarks, antiquarks or gluons; (2) replace all propagators
that do not lie between the photon vertices by special propagators;
(3) project out the good fields
and couple them with $t$-parton soft matrix elements.
The quark-mass terms follow in the same way, except that one (or more) of the
gluon legs that connect hard and soft parts is replaced by a spurion. In the
hard parts the partons are kept massless.
By the identification of the electromagnetic gauge invariant sets of hard scattering diagrams in the massive case, the unraveling of higher twist contributions is now completed. 

We thank P.J. Mulders for useful comments.
This work was supported by the Foundation for Fundamental Research on Matter
(FOM) and the National Organization for Scientific Research (NWO).

\vspace{0.2cm}
\vfill
{\small

\end{document}